\documentclass[usenatbib,galley]{mn2e}
\usepackage{graphicx}

\title{Inner disc rearrangement revealed by dramatic brightness variations in the young star PV~Cep}

\author[M. Kun et al.]{M. Kun$^{1}$\thanks{E-mail: kun@konkoly.hu}, E. Szegedi-Elek$^{1}$, 
A. Mo\'or$^{1}$, \'A. K\'osp\'al$^{2}$, P. \'Abrah\'am$^{1}$, 
\newauthor D. Apai$^{3}$, Z. T. Kiss$^{1}$, P. Klagyivik$^{1}$, T. Yu. Magakian$^{4}$, Gy. Mez\H{o}$^{1}$, 
\newauthor T. A. Movsessian$^{4}$, A. P\'al$^{1}$, M. R\'acz$^{1}$, J. Rogers$^{3}$ \\
$^{1}$Konkoly Observatory, Konkoly Thege ut 15-17, H-1121 Budapest, Hungary \\
$^{2}$Leiden Observatory, Leiden University, Leiden, The Netherlands \\
$^{3}$Space Telescope Science Institute, 3700 San Martin Drive, Baltimore, 
MD 21218, USA \\
$^{4}$V. A. Ambartsumian Byurakan Astrophysical Observatory, 0213 Aragatsotn prov., Armenia \\}
\begin{document} 


\maketitle

\label{firstpage}

\begin{abstract}  
Young Sun-like stars at the beginning of the pre-main sequence (PMS) evolution are 
surrounded by accretion discs and remnant protostellar envelopes. Photometric 
and spectroscopic variations of these stars are driven by interactions 
of the star with the disc. Time scales and wavelength dependence of the variability 
carry information on the physical mechanisms behind these interactions. 
We conducted multi-epoch, multi-wavelength study of PV~Cep, a strongly variable, 
accreting PMS star. By combining our own observations from 2004--2010 
with archival and literature data, we show that PV Cep started a 
spectacular fading in 2005, reaching an $I_\mathrm{C}$-band amplitude of 4~mag. 
Analysis of variation of the optical and infrared fluxes, colour indices, 
and emission line fluxes suggests that the photometric decline in 2005--2009 
resulted from an interplay between variable accretion and circumstellar extinction: 
since the central luminosity of the system is dominated by accretion, a modest drop 
in the accretion rate could induce the drastic restructuring of the inner disc. 
Dust condensation in the inner disc region might have resulted in the enhancement of 
the circumstellar extinction. 
\end{abstract}

                                                                               
\section{Introduction}
\label{Sect_1}

Eruptive young stars, defined by their episodically increased optical and infrared 
brightness, form a small subclass of Sun-like  PMS stars. 
Traditionally they are divided into two groups: FU Orionis type stars (FUors) exhibit
an initial brightening of 5~mag during several months or years, followed by a fading 
phase of up to several decades or a century. The other group, EX Lupi-type stars (EXors)
are characterized by relatively short, recurrent outbursts which last some weeks to months, 
and the time between the eruptions ranges from months to years. In both classes
the outbursts are believed to be powered by enhanced accretion from the circumstellar 
disc on to the star. In the standard picture the inward spiralling material piles up close 
to the inner edge of the accretion disk, and falls on to the stellar surface as a result 
of a gravitational and thermal instability \citep{Zhu}. 

In 2003, the outburst of the low-mass young stellar object V1647~Ori triggered a 
large number of  observations from X-ray to far-infrared wavelengths.	
Based on these ground-based and space-born measurements it was suggested that 
this object fits neither into the FUor nor EXor class. It was outlined that the outburst 
-- unlike in other young eruptive stars where enhanced accretion is the main 
physical driver behind the brightening -- consisted of the combination of two effects 
of comparable amplitude. They are an intrinsic brightening related to the appearance 
of a new, accretion-related hot component in the system, and a dust-clearing 
event which reduced the extinction along the line-of-sight \citep*[e.g.][]{AP07,ABR08}.
The simultaneity of the accretion and extinction changes suggests that they might
be physically linked, and the changing accretion luminosity causes changes in the 
inner disc structure. Another example for such combined processes may be V1184~Tau 
\citep{Alves97} whose optical and infrared photometric observations have shown 
that the large photometric decline was associated with variation in the inner 
disc structure \citep{Grinin09}. That the variable extinction may have an important 
role in the brightening refines the picture of the FUor/EXor phenomenon, moreover 
such rearrangements of the inner disc structure have a potentially high importance 
for the evolution of the terrestrial zone of circumstellar disk.

The target of our multi-wavelength variability study, PV~Cep is an eruptive young 
star, which was originally classified as member of the EXor class by \citet{Herbig89}, 
on the basis of its large outburst in 1977--1979 \citep{Cohen81}.
Currently the star undergoes significant brightness changes which we have been 
monitoring for years. Our first observations indicated that PV~Cep might be another 
example for accretion-induced structural changes, and this motivated our further
detailed study of the object. The role of variable circumstellar extinction was 
also apparent in the dramatic brightness decline observed in the near infrared 
between 2008 April and June by \citet{Lorenzetti09}. Several further observed properties
distinguish PV Cep from EXors, and indicate that the object may be
more similar in nature to V1647~Ori. Such properties  are e.g. the time-scale 
of the outburst ($\sim 2$\,years), the A-type absorption spectrum observed 
at the beginning of the outburst \citep{Cohen81}, the Class~I spectral 
energy distribution (SED) \citep{Connelley}, the associated high-mass 
circumstellar disc \citep[0.76\,M$_{\sun}$,][]{Hamidouche}, variable reflection 
nebula \citep{GM77,RNO}, molecular outflow \citep{Levreault84,AG02,Hamidouche}, 
and parsec-scale optical jet \citep*{Gomez97,Reipurth97,Neckel87}.

We carried out a comprehensive optical and infrared photometric 
and optical spectroscopic monitoring of PV~Cep during the low-brightness state 
in 2008--2010, and we have a few additional measurements from 2004, 2005, 
and 2006. In this paper we combine these data with archival and new near- and 
mid-infrared data, obtained in 2004--2010, to better understand the nature of PV~Cep
and the nature of physical processes which affect the inner disk structure and evolution.

\section{Observations}
\label{Sect_2}

\subsection{Photometry}
\label{Sect_phot}

\begin{figure}
\centering{\includegraphics[width=100mm]{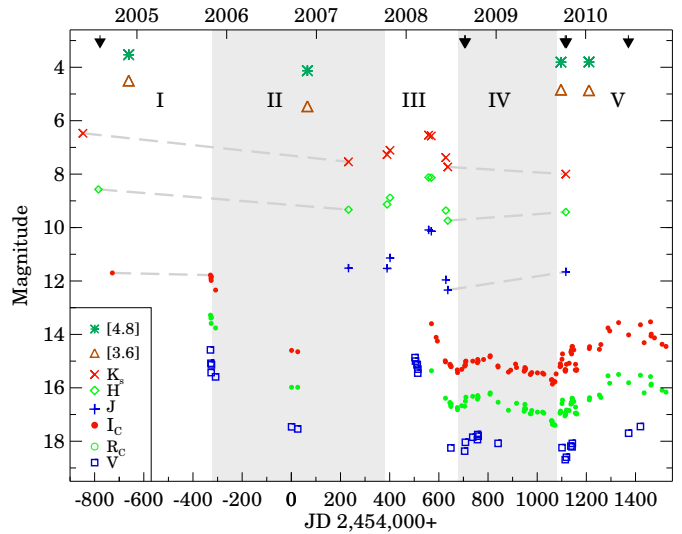}}
\caption{{\it V\/}, $R_\mathrm{C}$, $I_\mathrm{C}$, 
{\it J\/}, {\it H\/}, {\it K\/}$_\mathrm{s}$, and IRAC [3.6] and 
[4.8] light curves of PV~Cep between 2004 May and 2010 September. 
{\it V\/} magnitudes in 2008 February were taken from the {\it INTEGRAL} 
OMC Archive. Near-infrared data before 2009 are taken from 
\citet{Lorenzetti09} and \citet{Connelley}. The light grey dashed lines, spanning the 
long time-gaps without data, were drawn to clearly identify the coherent data sets. 
Arrows indicate the dates of our spectroscopic observations. Roman numbers and shading 
indicate the five characteristic segments of the light curves. }
\label{Fig_lcmulti_long}
\end{figure}

Photometric observations in the $VR_\mathrm{C}I_\mathrm{C}$ bands spanning 
the time interval between 2004 September 22 and 2010 November 23 were 
performed with six instruments on five telescopes. Most of the data were obtained 
with the 60/90/180~cm Schmidt telescope of the Konkoly Observatory, equipped with 
a Photometrics AT~200 camera and with the 1-m RCC telescope of the Konkoly 
Observatory, equipped with a Princeton Instruments VersArray:1300B camera.
PV Cep was also observed with the IAC--80 telescope of the
Teide Observatory (Spain) between 2009 October 25 and November 7.
More detailed description of these instruments is given in \citet{AP07}.
In 2008 August and 2009 October we obtained {\it VRI} images with the CAFOS 
instrument installed on the 2.2~m telescope of the Calar Alto Observatory (Spain).
Exposure times were between 90 and 180\,s per image.
The {\it I\/}$_\mathrm{C} = 11.70\pm0.04$ magnitude on 2004 September 22 was 
derived from a CCD-image obtained with the 2.6-m telescope of Byurakan Observatory, 
Armenia, and shown in the paper of \citet{MMSN08}. The image of PV~Cep 
was saturated, and its magnitude was derived by fitting the PSF to the 
unsaturated wing of the stellar image. On 2010 April 8 we used the Electro Multiplying (EM)  
Andor Technology iXon$^\mathrm{EM}+888$  camera at the RCC telescope of the 
Konkoly Observatory. We obtained a series of 100 exposures in the {\it I}$_\mathrm{C}$ band 
with exposure times of 20\,s per image in EM mode (the nominal EM gain was 50). 
The image processing was made with  the software tools written by 
\citet{Pal}. All the other images were reduced in {\sevensize  IRAF}.
In order to transform instrumental magnitudes into the standard system, 
we calibrated 16 stars in the field of view of the 1-m telescope 
($7\arcmin\times7\arcmin$) in $VR_\mathrm{C}I_\mathrm{C}$ bands. Calibration 
was made during six photometric nights. 
Standard stars in NGC~7790, published by \citet{Stetson}, were used as reference. 
Equatorial coordinates and derived $VR_\mathrm{C}I_\mathrm{C}$ magnitudes with 
uncertainties of the comparison stars are listed in Table~\ref{Tab_comp}.
The results of the photometry for PV~Cep are presented in Table~\ref{Tab_phot}
and plotted in Fig.~\ref{Fig_lcmulti_long}. 
The photometric errors were derived from quadratic sums of the formal errors 
of the instrumental magnitudes and those of the coefficients of the transformation 
equations. The telescope used for each observation is shown in the last column of
Table~\ref{Tab_phot}.

\begin{table*}
\begin{minipage}{126mm}
\begin{center}
\caption{Comparison stars$^*$}
\label{Tab_comp}
{\footnotesize
\begin{tabular}{lccccc}
\hline
N & RA(2000) & D(2000) & {\it V}\,($\Delta V$) & {\it R}$_\mathrm{C}$\,($\Delta R_\mathrm{C}$) & {\it I}$_\mathrm{C}$\,($\Delta I_\mathrm{C}$)  \\
 &  h m s & d \ \arcmin \ \arcsec & [mag] & [mag] & [mag] \\
\hline
C1  & 20 46 28.89 & +67 59 05.4 & 14.072\,(0.012) &  13.595\,(0.012) &  13.200\,(0.008) \\
C2  & 20 46 26.50 & +67 58 11.8 & 14.608\,(0.005) &  13.948\,(0.004) &  13.495\,(0.005) \\
C3  & 20 46 10.44 & +67 55 44.2 & 15.134\,(0.005) &  14.205\,(0.004) &  13.637\,(0.004) \\
\hline
\end{tabular}}
\end{center}
\medskip
{\footnotesize $^*$The full Table is available as Supporting Information with
the online version of this article.}
\end{minipage}
\end{table*}

\begin{table*}
\begin{minipage}{126mm}
\begin{center}
\caption{Results of the photometry of PV Cep$^**$}
\label{Tab_phot}
{\footnotesize
\begin{tabular}{rccccl}
\hline
JD & Date &{\it V}\,($\Delta V$) &  {\it R}$_\mathrm{C}$\,($\Delta R_\mathrm{C}$) & {\it I}$_\mathrm{C}$\,($\Delta I_\mathrm{C}$) & Telescope \\
(2454000+) & yyyymmdd & [mag] & [mag] & [mag]  \\
\hline
$-$329 & 20051027  & 14.58\,(0.16) &  13.28\,(0.06)  & 11.78\,(0.06) & RCC \\
$-$327 & 20051029  & 15.08\,(0.04) &  13.39\,(0.04)  & 11.90\,(0.11) & RCC  \\
$-$326 & 20051030  & 15.43\,(0.06) &  13.59\,(0.04)  & 11.98\,(0.06) & RCC  \\
$-$325 & 20051031  & 15.14\,(0.07) &  13.36\,(0.01)  & 11.84\,(0.16) & RCC  \\
\hline			 		 	 
\end{tabular}}		 
\end{center}
\medskip		 
{\footnotesize $^{**}$The full Table is available as Supporting Information with
the online version of this article.}
\end{minipage}		 
\end{table*}		 

\subsection{Near-infrared observations}
\label{Sect_nir}

Near-infrared, {\it JHK\/}$_\mathrm{s}$  
measurements of PV~Cep were obtained using the NIC-FPS camera on the 
ARC~3.5-m Telescope at Apache Point Observatory on 2009 October 12 (JD 2455116.5).
Dark and domeflat images were also obtained, and four-point dithering was applied
to ensure proper  sky subtraction. To avoid saturation, the {\it K\/}$_s$ images
were slightly defocussed, and, to exclude a nearby bright star from
the field of view during the {\it H\/} and {\it K\/}$_s$ exposures, 
the camera was rotated. The images were reduced in {\sevensize IRAF}. 
Magnitudes were derived by aperture photometry, 
and were transformed into the standard system by comparing them
with the 2MASS magnitudes  \citep{2MASS} of four stars in the image field. These stars are 
as follows: 2MASS 20462889+6759054 (C1 in Table~\ref{Tab_comp}), 
2MASS 20462650+6758118 (C2), 2MASS 20461044+6755442 (C3), and  
2MASS 20461128+6800012. The derived magnitudes are as follows: 
{\it J\/}$=11.66\pm0.04$,  {\it H\/}$= 9.42\pm0.04$,  {\it K\/}$_s= 8.00\pm0.10$.
		 
\subsection{Spectroscopy}
\label{Sect_spec}	

Intermediate resolution spectra of PV~Cep were obtained on 2008 August 26 and 29,
2009 October 9, 11, and 14, using the CAFOS instrument on the
2.2-m telescope of the Calar Alto Observatory. The R--100 grism covered the 
5800--9000\,\AA\ wavelength range. The spectral resolution of CAFOS, using a 
1.5-arcsec slit, was $\lambda / \Delta \lambda \approx 3500$ at $\lambda=6600$\,\AA. 
The spectrum of a He--Ne--Rb lamp was regularly observed for wavelength calibration.  
Broadband $VR_\mathrm{C}I_\mathrm{C}$ photometric images were taken immediately before 
the spectroscopic exposures for flux calibration.
We reduced and analysed the spectra using standard {\sevensize  IRAF} routines.
A further R--100 spectrum, obtained on 2004 August 7, and shown in fig.~2 of 
\citet{Kun09}, is also available.

\subsection{\textit{Spitzer} data}
\label{Sect_spitzer}
%
We observed PV~Cep using the \textit{Spitzer Space Telescope} 
between 2004 October and 2005 August with the IRAC, IRS and MIPS
instruments. In addition, we used further archival data from IRAC and MIPS. 
Moreover, we conducted a monitoring with the IRAC instrument
during the post-Helium phase in 2009 September and 2010
January. A log of all Spitzer observations of PV~Cep is presented in
Table~\ref{tab_spitzerlog}.  The data reduction procedures are described in 
Appendix~A, available as Supporting Information with
the online version of this article. The results of Spitzer photometry are 
shown in Table~\ref{Tab_Spitzer}.

\begin{table*}
\begin{minipage}{126mm}
\caption{Log of \textit{Spitzer} observations of PV Cep}
\label{tab_spitzerlog}
\centering
{\small
\begin{tabular}{c c c c c}
\hline
\hline
Instrument   & Wavelength                 & Date                 & AOR                  & Program \\
\hline
IRAC         & 3.6, 4.5, 5.7, 8.0$\,\mu$m & 2004 Oct 29          & 11571712             & 3716  \\
IRS          & 5.2 -- 37.2$\,\mu$m        & 2004 Oct 23          & 12287488             & 3716  \\
MIPS SED     & 55 -- 95$\,\mu$m           & 2005 Aug 03          & 11571968             & 3716  \\
IRAC         & 3.6, 4.5, 5.7, 8.0$\,\mu$m & 2006 Nov 26          & 18955008             & 30760 \\
MIPS         & 24, 70$\,\mu$m             & 2007 Feb 26          & 19962624, 19962880   & 30574 \\
IRAC post-He & 3.6, 4.5$\,\mu$m           & 2009 Sep -- 2010 Jan & 35591168 -- 35678720 & 60167 \\
\hline
\end{tabular}}
\end{minipage}
\end{table*}

\begin{table*} 
\begin{minipage}{126mm}
\caption{\textit{Spitzer} photometry for PV Cep. All fluxes are color-corrected and given in Jy$^\dagger$.}
\label{Tab_Spitzer}
\begin{center}
{\small
\begin{tabular}{c c c c c c c}
\hline
\hline
Date        & F$_{3.6\mu\rm m}$ & F$_{4.5\mu\rm m}$ & F$_{5.7\mu\rm m}$ & F$_{8.0\mu\rm m}$ & F$_{24\mu\rm m}$ & F$_{70\mu\rm m}$ \\
\hline
20041029 & 4.47\,(0.09) & 6.98\,(0.14) & 8.34\,(0.17) & 11.85\,(0.23) &             &            \\
20061126 & 1.84\,(0.04) & 3.99\,(0.08) & 5.45\,(0.11) & 8.16\,(0.17)  &             &            \\
20070226 &            &            &            &             & 27.12\,(1.04) & 35.22\,(2.13)\\
20090916 & 3.54\,(0.03) & 5.97\,(0.02) \\
\hline
\end{tabular}}
\end{center}
\medskip		 
{\footnotesize $^\dagger$The full Table is available as Supporting Information with
the online version of this article.}
\end{minipage}
\end{table*}

\section{Results}
\label{Sect_3}

\subsection{Photometric variations}
\label{Sect_lc}

\paragraph*{Light curves of PV~Cep in 2004--2010.} 
The light curves of PV~Cep in the {\it V\/}, $R_\mathrm{C}$, $I_\mathrm{C}$, 
{\it J\/}, {\it H\/}, {\it K\/}$_\mathrm{s}$, and IRAC 3.6 and 4.5\,$\mu$m bands 
between 2004 May and 2010 November are shown in Fig.~\ref{Fig_lcmulti_long}. 
Based on the shape of the $I_\mathrm{C}$ curve we divided the covered period 
into five segments, indicated by shading and Roman numbers. The first segment, thereinafter 
referred to as the {\it bright state\/}, is the interval between 2004 July and 
2005 October. Segment~II, the {\it fading phase}, is the period between 2005 October 
and 2007 October, when the star faded significantly in each photometric band. 
Segment~III, the {\it transient peak\/}, covers the brightening between 2007 October 
and 2008 February, and the subsequent sharp decline until 2008 June. 
Segment~IV is the {\it low-brightness state\/} between 2008 June and 2009 August, 
when the optical brightness slowly decreased further, reaching a minimum of 
{\it I}$_\mathrm{C} = 15.86$ on 2009 August 18. After this date, in segment~V, 
the optical and infrared fluxes, except the {\it K\/}$_\mathrm{s}$-band flux, 
started rising, and large, short-term brightness fluctuations appeared in the 
optical light curves. We refer to this interval as the {\it rising phase\/}. 
The total amplitude of the $I_\mathrm{C}$ light curve is about 
4~magnitudes over the period 2005 November and 2009 September. 

\begin{figure}
\resizebox{\hsize}{!}{\includegraphics{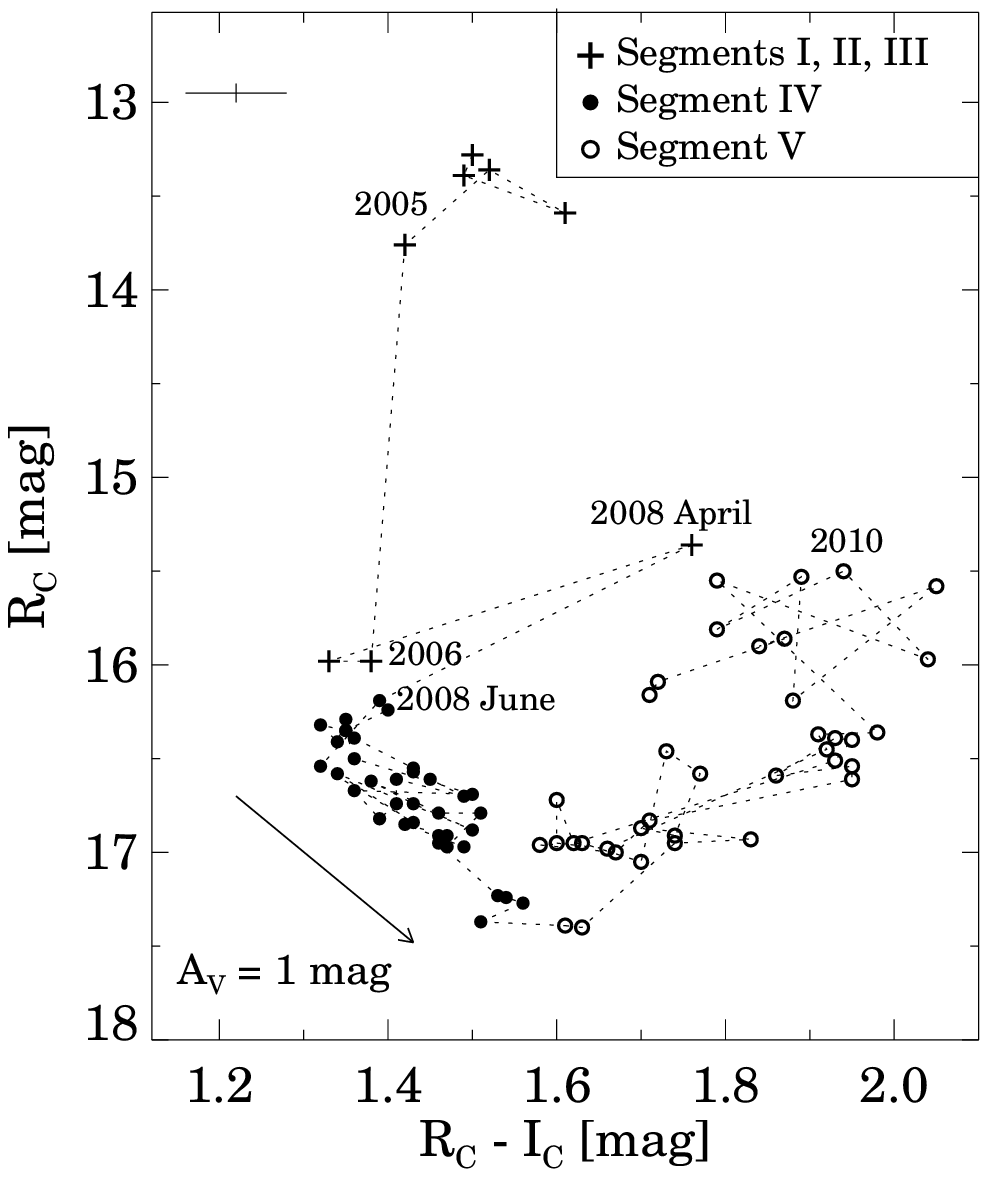}\includegraphics{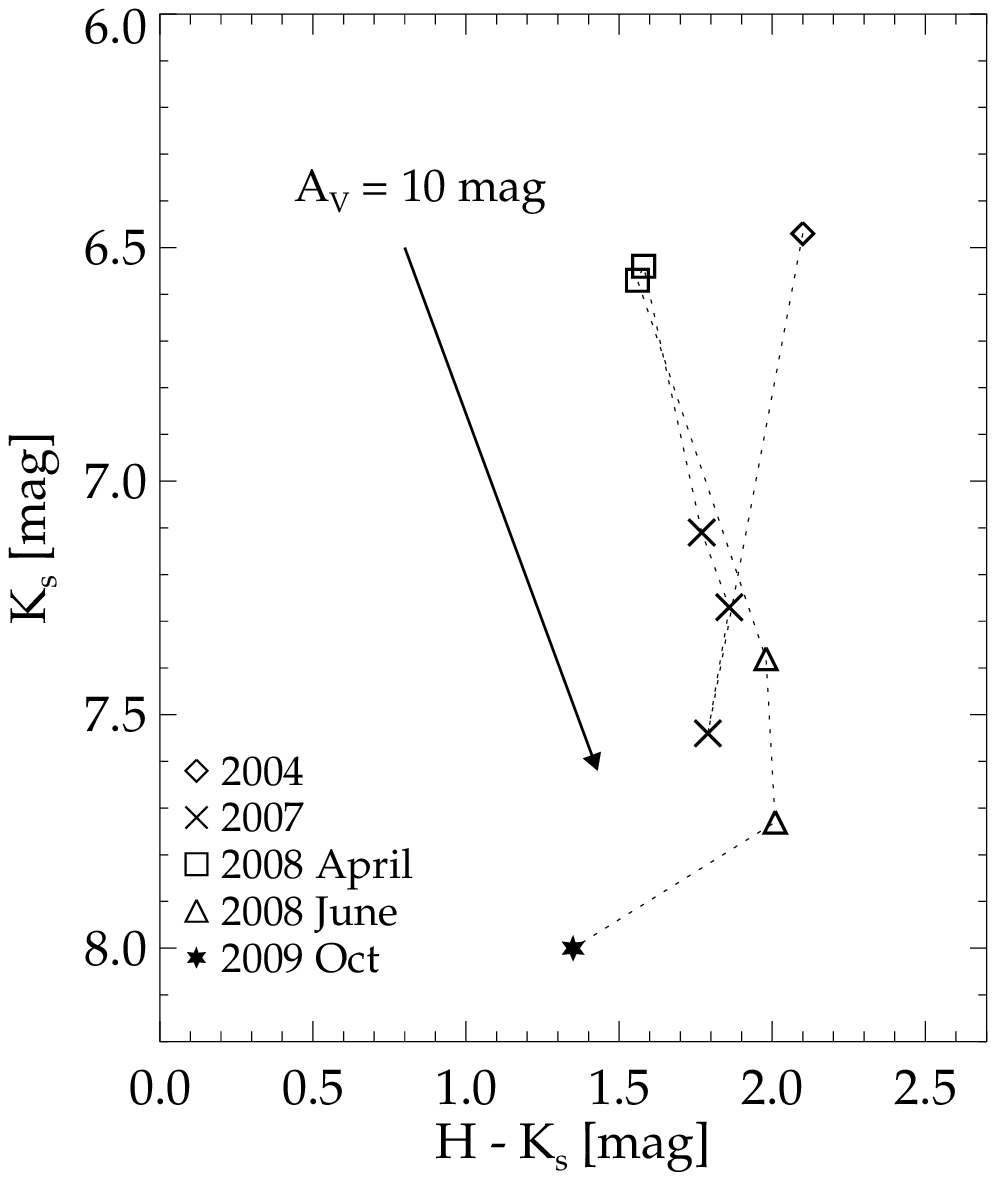}}
\caption{{\it Left}: $R_\mathrm{C}$ magnitude as a function of the $R_\mathrm{C} - I_\mathrm{C}$ 
colour index. Consecutive data points are connected by dotted lines. The arrow indicates
the colour dependence of the interstellar extinction. 
The typical error bars are shown in the upper left corner. {\it Right}: {\it K\/}$_\mathrm{s}$ 
magnitude as a function of the $H-K_\mathrm{s}$ colour index based on all data obtained
between 2004 and 2010 \citep{Connelley,Lorenzetti09}. Consecutive data points are connected.}
\label{Fig_cmd}
\end{figure}

\paragraph*{Colour--magnitude diagrams.}
In order to get insight into the origin of the observed variations 
we plotted colour--magnitude diagrams. Figure~\ref{Fig_cmd} (left) shows  that the fading in 
2005--2006 was nearly grey in the {\it R}$_\mathrm{C}$ and {\it I}$_\mathrm{C}$  bands.
Its reason may be that in minimum, whatever is the origin of the decline, the star itself 
is too faint to be detected, only the light scattered from the disc atmosphere, 
thus bluer in colour, can be observed. The transient peak shows different colour 
variation: the star became redder and brighter and then bluer and fainter. 
This colour behaviour is characteristic of UX Orionis 
(UXor) type variables  close to the light curve minima \citep[e.g.][]{Bibo}, 
and is attributed to the increasing proportion of scattered light when the star 
is obscured by a circumstellar dust clump. This diagram suggests that the optical 
source in the dim phases (segments II and IV of the light curve) was the starlight 
scattered from the disc atmosphere. During the low-brightness phase (filled circles) 
PV~Cep was redder when fainter: this behaviour can result from either enhanced 
extinction or vanishing hot spots on the stellar surface due to the decreasing 
accretion rate. In the rising phase the star turned brighter and redder, 
indicating that instead of the scattered light direct light from the central 
star could be detected. The data obtained in 2010 indicate that the star 
completed a full loop in this diagram from 2008 April to 2010 April. 

The near-infrared data in Fig.~\ref{Fig_cmd} (right) show that the fading in 2005--2006 
was accompanied by a slight decrease of the $H-K_\mathrm{s}$ colour index, indicative of
decreasing emission from the dust sublimation zone of the disc, whereas an extinction 
change of $A_\mathrm{V} \approx 7$~mag can account for the photometric variations around the 
transient peak in 2008. The different colour behaviour suggests the different nature of the 
two brightness drops. The strong decrease of the $H-K_\mathrm{s}$ colour index at the 
bottom of the diagram, measured in 2009 October, clearly suggests the fall of the inner 
disc emission during the low-brightness phase. 

\begin{figure}
\resizebox{\hsize}{!}
{\includegraphics{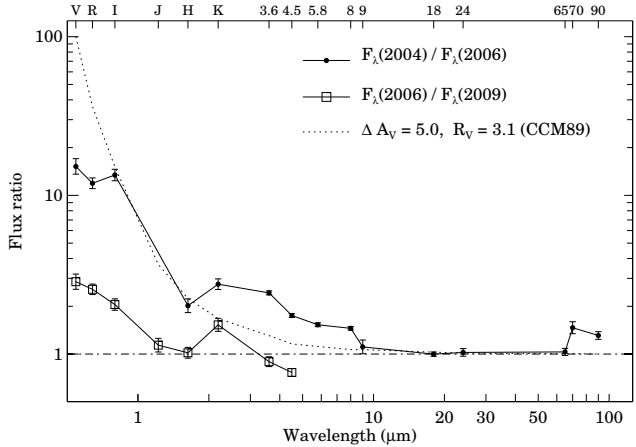}}
\caption{Flux changes of PV Cep between the different segments of the light curve, as 
a function of the wavelength. The dotted line shows the flux ratios corresponding to an 
increase of the extinction by $A_\mathrm{V} = 5$~mag  \citep[][asuming $R_\mathrm{V}=3.1$]{CCM89}.}
\label{Fig_deltaft}
\end{figure}

\paragraph*{Variations in the mid- and far infrared} 
The \textit{Spitzer} data obtained in the bright, fading, and rising phases of the 
light curve (Table~\ref{Tab_Spitzer}) allow us to inspect the wavelength dependence of 
variations in the mid- and far-infrared regions. We plotted in Fig.~\ref{Fig_deltaft} the 
flux ratios between the different segments of the light curve as a function of wavelength.
In addition to our own measurements, the bright state data also include {\it H\/} and 
$K_\mathrm{s}$ data from \citet{Connelley}, and the fading phase data set contains 
{\it JHK}$_\mathrm{s}$ data measured on 2007 May 13 \citep{Lorenzetti09}, as well as 
the \textit{AKARI IRC\/} \citep{Ishihara} fluxes at 9 and 18\,$\mu$m, and \textit{AKARI FIS\/} 
catalog data \citep{Yamamura} at 65, 90, 140, and 160\,$\mu$m, each obtained 
between 2006 May 8 and 2007 August 26. The wavelength dependence of variations is 
obviously different from that of the dust extinction: the excess emission in 
the bright state  with respect to fading phase had an optical, a near- and mid-infrared, 
and a far-infrared component, indicative of changing emission from both the central star 
and various parts of the disc. The optical and near-infrared excess components are separated 
by a minimum in the {\it H\/} band. The wavelength interval 10--65\,$\mu$m was 
unaffected by the fading, while the system also dimmed in the 70--90\,$\mu$m 
region. The $F_{\lambda}$(2006)/$F_{\lambda}$(2009) curve shows that both the star and 
the inner disc continued fading after the transient peak. 

\subsection{Variations in the nebula RNO~125}
\label{Sect_nebula}

\begin{figure}
\resizebox{\hsize}{!}{\includegraphics{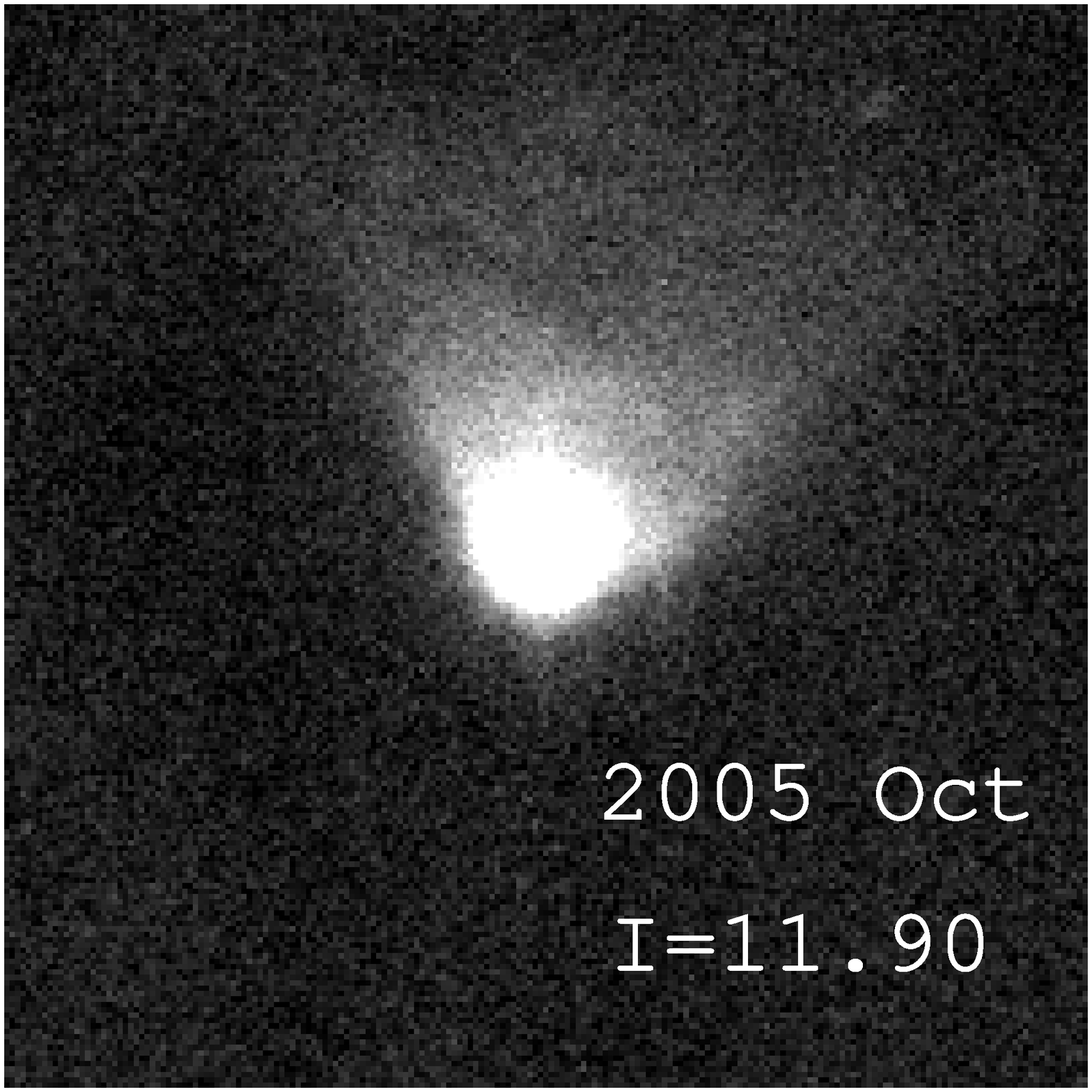}\includegraphics{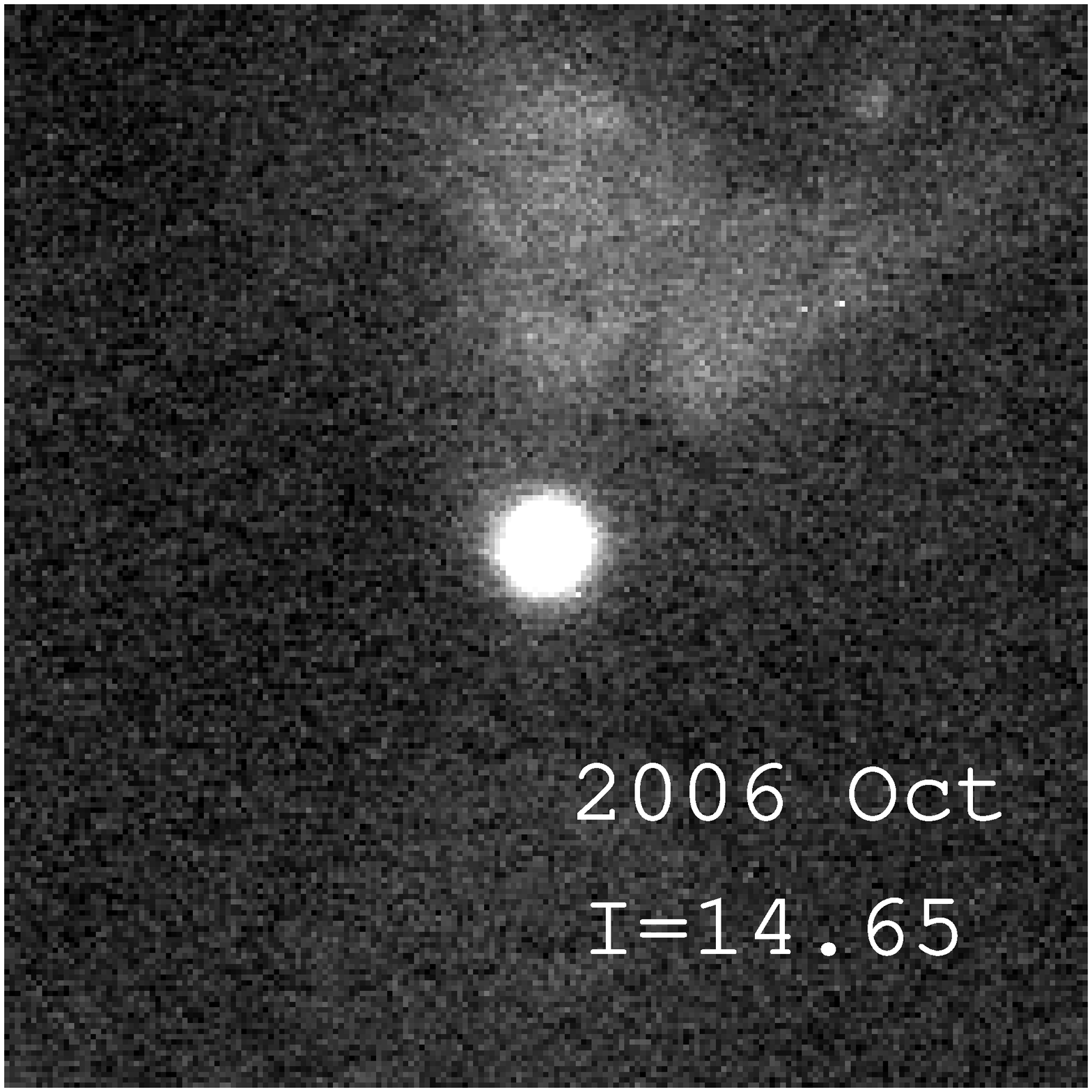}}
\resizebox{\hsize}{!}{\includegraphics{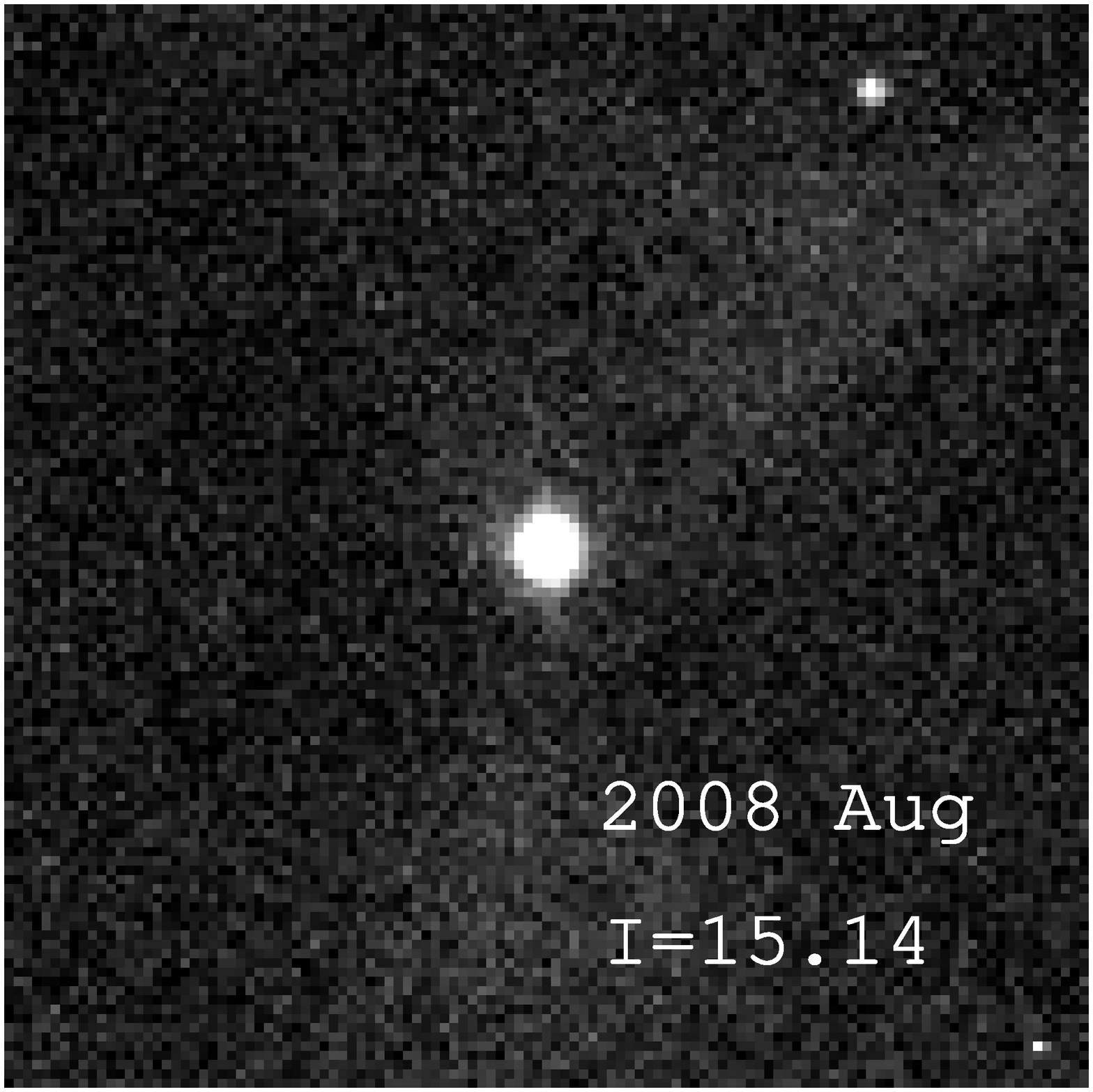}\includegraphics{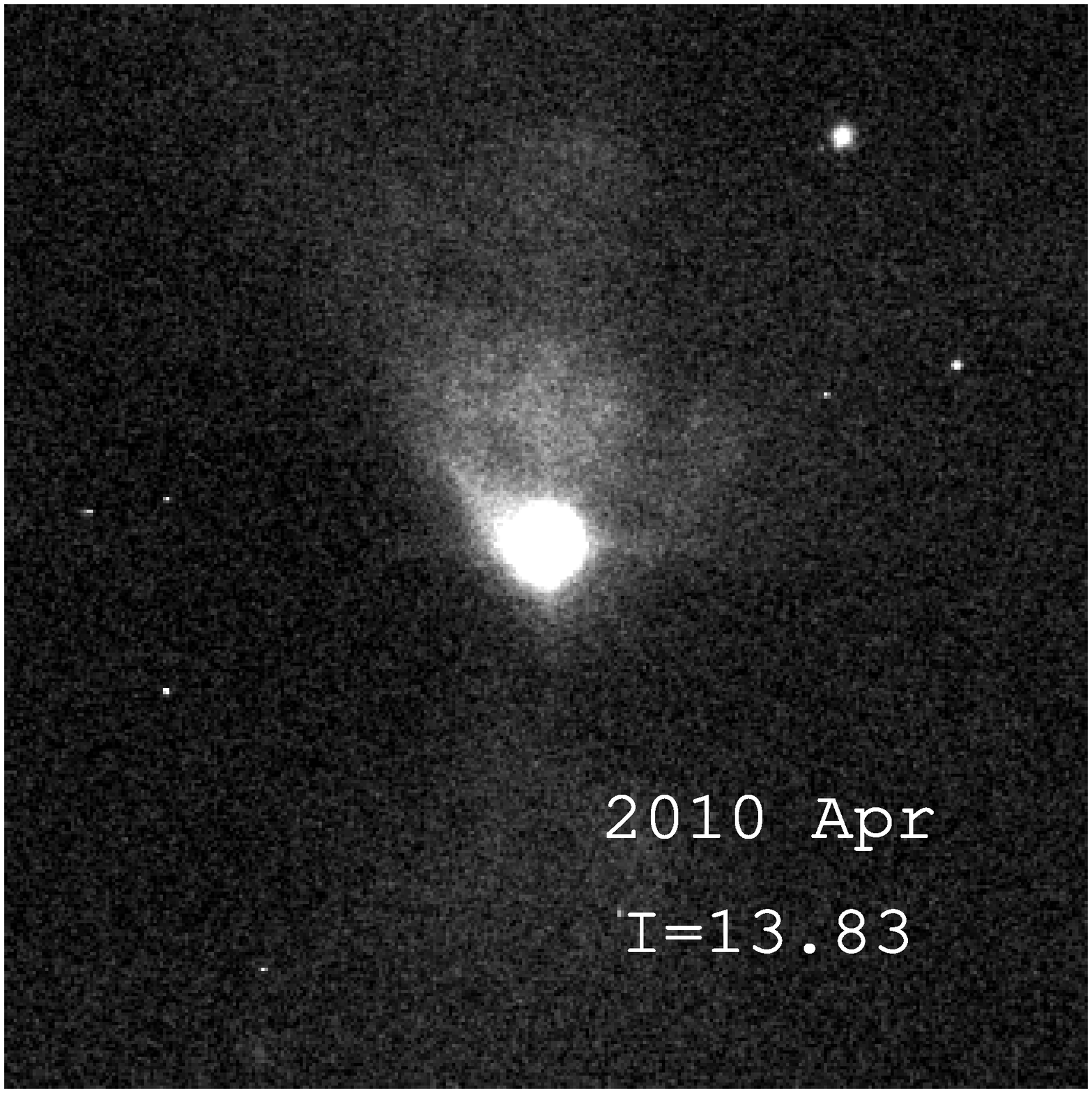}}
\caption{I-band images of the nebula RNO~125 at different dates. The size of each image is
$1\arcmin \times 1\arcmin$. North is up, and east is to the right. }
\label{Fig_neb}
\end{figure}

\paragraph*{Variations in the nebula RNO~125.} 
RNO~125 was fan-shaped and bright in the bright state \citep{MMSN08}, and the bright 
head of the cometary nebula was centered on the star (Fig.~\ref{Fig_neb}, upper left). 
A dark band appeared between the star and the nebula during the fading phase 
(Fig.~\ref{Fig_neb}, upper right). The nebula quickly faded and exhibited a strongly 
changing shape after the transient peak. Similar phenomenon was observed in 1977--1979 
\citep[see fig.~1 of ][]{Cohen81}. The nebula was barely visible in our images during 
the low-brightness phase and at the beginning of the rising phase (Fig.~\ref{Fig_neb}, 
lower left). The head of the nebula started brightening in 2010 (Fig.~\ref{Fig_neb},
lower right).

The variable shape of the nebula suggests that the illumination changed 
not only due to the fading of the star, but also due to the changing geometry of 
the dust distribution close to the star, resulting in rapidly changing shadows on 
the matter inside the outflow cavity. The rebrightening of the nebula  started close 
to the star: those parts appeared first which dimmed when the star 
started declining. 


\subsection{Variations in the optical spectrum}
\label{Sect_spvar}

No photospheric absorption lines can be identified in the spectra. 
With the exception of H$\alpha$, the strong emission lines are symmetric 
at our resolution. The forbidden lines are systematically blueshifted with respect to the 
permitted lines by some 100--200\,km\,s$^{-1}$, indicating their origin from the approaching 
lobe of a wind.
Table~\ref{Tab_lines} shows the equivalent widths and fluxes of the strongest 
emission lines in the optical spectra of PV~Cep, obtained in the bright, low-state, and
rising phases of the light curve. The fluxes were determined using the $R_\mathrm{C}$ 
and $I_\mathrm{C}$ magnitudes, measured simultaneously with the spectroscopic 
observations, and the fluxes corresponding to zero magnitude tabulated by \citet{Glass}. 
Since only $I_\mathrm{C}$ magnitude is available for 2004, we estimated the 
$R_\mathrm{C}$ magnitude in 2004 using the average colour index 
$R_\mathrm{C}-I_\mathrm{C} = 1.50$, measured at the end of the bright state in 2005. 
The  variations of the emission lines can be summarized as follows. 
(1) The equivalent width of the H$\alpha$ line remained nearly unchanged during the 
photometric decline, that is, the flux of the H$\alpha$ emission and 
the continuum flux underwent equal fading, suggesting that the star and the 
H$\alpha$ emitting region were attenuated by a circumstellar dust structure.
(2) The fluxes of the forbidden lines decreased only slightly during the 
decline of the stellar flux (the equivalent widths strongly increased), indicating that the 
formation regions of these lines were unaffected by the changes leading to the fading. 
(3) The decline of  the Ca\,{\sevensize II} line fluxes was significantly larger 
than that of the underlying continuum: their equivalent widths were halved between 
2004 and 2008. Since these lines are accretion tracers \citep*{Muzerolle98}, the accretion 
rate of PV~Cep declined between 2004 August and 2008 August. 

\begin{table}
\caption{Equivalent widths and fluxes of emission lines}
\label{Tab_lines}
{\footnotesize
\begin{tabular}{lc@{\hskip1mm}c@{\hskip1mm}c@{\hskip1mm}cc@{\hskip1mm}c@{\hskip1mm}c@{\hskip1mm}c}
\hline         
Id. &  \multicolumn{4}{c}{EW (\AA)} & \multicolumn{4}{c}{F($10^{-17}$\,Wm$^{-2}$)}  \\
& 2004 & 2008 & 2009 & 2010 & 2004 & 2008 & 2009 & 2010 \\
\hline
H$\alpha$                    & 104 & 132 & 100 & 90   & 99  & 6.6 & 3.8 & 9.3  \\
Ca{\sevensize  II} 8498      & 77  & 42  & 62  & 39.5 & 168 & 4.1 &  6.3 & 11.0  \\ 
Ca{\sevensize  II} 8542      & 74  & 39  & 59  & 37.4 & 161 & 3.8 &  5.9 & 10.4  \\
Ca{\sevensize  II} 8662      & 64  & 32  & 51  & 30.0 & 139 & 3.1 &  5.1 & 8.3  \\
$[$O{\sevensize I}$]$ 6300   & 8.1 & 58  & 82  & 37.0 & 4.8 & 2.9 &  3.1 & 3.8  \\
$[$O{\sevensize I}$]$ 6363   & 1.7 & 20  & 27  & 13 & 1.6 & 1.0 &  1.0   & 1.3  \\
$[$S{\sevensize  II}$]$ 6717 & 1.2 & 4.6 & 9.2 & 3.3 & 1.1 & 0.23 & 0.35 & 0.34  \\
$[$S{\sevensize  II}$]$ 6731 & 1.6 & 8.6 & 15.6& 5.9 & 1.5 & 0.43 & 0.59 & 0.61  \\
\hline
\end{tabular}}
\end{table}

\subsection{Physical properties of PV~Cep}
\label{Sect_phys.params}

\paragraph*{Mass and luminosity.} Given that the amplitude of the photometric variations 
was smallest in the {\it H} band (Fig.~\ref{Fig_deltaft}), we estimated the luminosity 
of PV~Cep from the {\it H} magnitude measured in the low-brightness state on 2008 June 18, 
assuming that the total flux in the {\it H} band originated from the photosphere. 
We adopted a distance of 325~pc for PV~Cep \citep{SCKM}, an effective temperature 
$T_\mathrm{eff}= 5500$~K, corresponding to the spectral type G8--K0 found by 
\citet{MM01}\footnote{The widely accepted spectral type of PV~Cep, 
A5 \citep{The94}, is based on the Balmer absorption lines, observed  
in the spectrum at the beginning of the outburst in 1977 \citep{Cohen81}. 
No spectrum, obtained after the outburst, exhibited Balmer absorption lines, which thus
might have been originated from the hot inner disc in early outburst phases, even as 
in the case of V1057 Cyg and V1647 Ori \citep[cf.][]{Welin71,Briceno04}.}, a total 
extinction $A_\mathrm{V}=12.0$~mag, obtained by dereddening the near-infrared colour 
indices  \citep{Lorenzetti09} on to the T~Tauri locus of the $J-H$ vs. 
$H-K_\mathrm{s}$ diagram \citep*{Meyer97}, and applied the bolometric correction
tabulated by \citet*{Hartigan} and unreddened colour index $V-H=1.69$ for the spectral 
type G8 \citep{KH95}. We obtained $L_{*} \approx 17$\,$L_{\sun}$,  
and a stellar radius $R_{*} \approx 4.0$\,R$_{\sun}$. Plotting the results on the HRD, 
together with the evolutionary models of \citet{PS99}, results in a mass of 
2.4\,M$_{\sun}$ and age of $< 1$\,Myr. 

\paragraph*{Disc accretion rates.} We estimated the low-brightness-state disc accretion 
rate of PV~Cep from the luminosity of the Ca\,{\sevensize II}~8542 line, measured 
in 2008 August (Table~\ref{Tab_lines}), using the empirical formula of \citet{Dahm08}. 
We obtain $L_\mathrm{acc}^\mathrm{low} \approx 41~L_{\sun}$,  
and  $\dot{M}_\mathrm{acc}^\mathrm{low} \approx 2.6 \times10^{-6}$\,M$_{\sun}$/yr. 
The corresponding values for the bright state, estimated from the flux change of 
the Ca\,{\sevensize II}~8542 line, are about twice higher.
The results show that the accretion rate of PV~Cep is significantly higher than  
typical PMS accretion rates \citep[$\sim 10^{-8}$\,M$_{\sun}$/yr, e.g.][]{Gullbring,Dahm08} 
both in the high- and low-brightness states, and the accretion luminosity is higher than
the photospheric luminosity. 

\section{Possible nature of the variations}
\label{Sect_scen}

{\it The decreased fluxes of the Ca\,{\sevensize II} emission lines of PV~Cep} between 
2004 and 2008 suggest that the fading was associated with decreasing accretion rate. 
The factor of two drop of the accretion luminosity, derived from the 
Ca\,{\sevensize II}~8542 lines, however, can account for only $\la~1$~mag variation in
the optical region. To explain the whole amplitude of $\sim~4$~mag, enhanced circumstellar 
extinction also has to be invoked. The excess extinction could arise from the
changes in the inner disc structure. Since $L_\mathrm{acc} > L_*$, the modest drop 
of the accretion rate substantially altered the total central luminosity, 
and the decreased central luminosity made the dust sublimation radius shrink, that is, 
a large amount of dust condensed in the inner disc region during the fading and 
the subsequent low-brightness  state. The bright-state luminosity of 
$L_\mathrm{acc}+L_{*} \approx 100~L_{\sun}$ implies $R_\mathrm{subl}\sim$\,0.7~AU 
\citep[cf. fig. 7 of ][]{Dullemond}, which might have shrunk to $\sim 0.4$~AU due to a 
factor of two drop in $L_\mathrm{acc}$. The observed drop in the $K_\mathrm{s}$-band flux 
between 2004 and 2007 suggests similar decrease in the dust sublimation radius,
if we assume that most of this flux is emitted from the dust sublimation
zone. Assuming normal interstellar gas-to-dust ratio, an excess extinction 
of $A_\mathrm{V}\sim 3$~mag requires a gas column density of 
$\sim~6\times10^{21}$\,cm$^{-2}$, which, taking into account 
the size of the dust condensation region, corresponds to a volume density of some 
$10^{9}$\,cm$^{-3}$. Considering typical midplane gas densities and scaleheights 
at the dust rim \citep{Dullemond}, the inclination of 62\degr\ and the fact 
that the disc of PV~Cep is at least 10 times more massive than a typical HAe star 
disc \citep{Hamidouche}, such densities along the line of sight are likely, and thus 
the newly formed dusty region may account for most of the observed optical and 
near-infrared fading of the star. The observed drop in the mid- and far-infrared fluxes 
indicates that parts of the optically thin outer flared disc atmosphere got into shadow. 

{\it The transient peak in 2008\/}, according to the colour--magnitude diagrams, resulted 
from variable obscuration along the line of sight. A very similar transient peak 
both in amplitude and time-scale appeared in the light curve of PV~Cep 
in 1979, following the end of the outburst \citep[fig. 2 of][]{Cohen81}, which suggests
that the temporary dust-clearing might have been related to the end of the outburst. 
A short-lived outflow, similar to that observed in V1647~Ori after the end 
of its outburst in 2006, and probably associated with the rearrangement of the stellar
magnetic field in response to the dropped accretion rate \citep{Brittain10} might have
blown out some dust from the line of sight. 

{\it The slowly descending light curve accompanied by reddening between 2008 June 
and 2009 August\/} may result from the dust condensation process.
The duration of the low-brightness period suggests that the dust, emerging in 
the line of sight, was not confined to dense clumps or warps in the disc, but rather 
it was associated with a restructuring of the inner disc due to the changing 
central luminosity.

{\it The brightening  accompanied by reddening in 2009--2010\/} may indicate that
the dust, condensed during the previous years, started evaporating. 
The increased Ca{\sevensize II} fluxes suggest that invigorated accretion 
may play role in the processes. The time-scales of the large-amplitude photometric 
fluctuations suggest that the inner dust rim is involved. Notably, two prominent 
peaks of the $I_\mathrm{C}$ light curve (around JD\,2,455,103 and JD\,2,455,143) 
are separated by 40 days, corresponding to the Kepler-period at a radius of 0.4\,AU.
The low $K_\mathrm{s}$-band flux, rising fluxes at 3.6 and 4.5\,$\mu$m, as well as 
the reappearance of the nebula suggest that the inner dust rim of the disc 
became less efficient in shadowing the outer parts of the circumstellar matter. 

This scenario can qualitatively explain the fading in the optical, 
near-, mid-, and far-infrared wavelength regions, and the simultaneous 
variations of the nebula and the Ca{\sevensize II} emission line fluxes. 
Our results suggest that PV~Cep differs from both the EXor type stars, characterized 
by outbursts driven by strongly enhanced accretion, and UXor stars, whose variability 
is caused by orbiting circumstellar dust structures. The high disc mass and 
persistent high accretion rate suggest that PV~Cep is a protostar-like object: 
most of its luminosity originates from accretion, and its large-scale 
photometric variations reflect the changing inner disc structure, resulted from 
unsteady accretion. Variations in the accretion rate may be triggered by gravitational 
instabilities in the massive disc of PV~Cep.

\section*{Acknowledgements}

Our results are partly based on observations obtained at the Centro 
Astron\'omico Hispano Alem\'an (CAHA) at Calar Alto, operated jointly by the 
Max-Planck Institut f\"ur Astronomie and the Instituto de Astrof\'{\i}sica 
de Andaluc\'{\i}a (CSIC). Our observations were supported by the
OPTICON. OPTICON has received research funding from the European
Community's Sixth Framework Programme under contract number
RII3-CT-001566. The 0.82-m IAC-80 Telescope is operated on the island of 
Tenerife by the Instituto de Astrof\'{\i}sica de Canarias in the Spanish 
Observatorio del Teide. This work makes use of observations made with the 
\textit{Spitzer Space Telescope}, which is operated by the Jet Propulsion 
Laboratory, California Institute of Technology under a contract with NASA. 
This publication makes use of data products from the Two Micron All Sky 
Survey, which is a joint project of the University of Massachusetts 
and the Infrared Processing and Analysis Center/California 
Institute of Technology, funded by the National Aeronautics and Space 
Administration and the National Science Foundation. This paper utilized data from 
the OMC Archive at LAEFF, pre-processed by ISDC. Financial support from the 
Hungarian OTKA grant K81966 is acknowledged. The research of \'A.K. is supported 
by the Nederlands Organization for Scientific Research.

\section*{Online Supporting Material}

\appendix

\section{\textit{Spitzer} data reduction}
\label{sect_spred}

\paragraph*{IRAC.} The IRAC observations of PV~Cep were performed in two
different observing modes. The observations in 2004 and those in
the framework of the post-Helium monitoring programme were performed
using the subarray readout mode. The observations in 2006 were performed
in high dynamic range readout mode with 0.6\,s and 12\,s integration
time.

The subarray mode allows the utilization of exposure time as short as
0.02\,s enabling the observation of bright sources that would saturate
the detector using the other readout modes. The data processing  
was started from the \textit{Spitzer} Science Center (SSC) basic calibrated
data (BCD) produced by the pipeline version S18.7. The BCD image cubes
of the 64 frames were combined into two-dimensional images using the
``irac-subcube-collapse'' {\sevensize IDL} routine provided by the SSC. The
observation performed in 2004 was made of nine different dither
positions and repeated three times resulting in 27 images at each
wavelength.  In the case of monitoring observations four dithering steps
were utilized, leading to four images at each wavelength. We
performed aperture photometry on the final images at each wavelength
using a modified version of the {\sevensize IDLPHOT} routines. The aperture radius
was set to 3 pixels, the sky background was computed in an annulus
with an inner radius of 3 pixels and a width of 4 pixels. In the
course of the sky estimates we used an iterative sigma-clipping
method, where the clipping threshold was set to 3$\sigma$. Following
the outline of \citet{Hora08} we also applied an array-dependent
photometric correction and a pixel-phase correction to the measured
flux densities. An aperture correction was then performed using the
values published in the IRAC Data Handbook. The individual flux
density values measured in the different dither positions (at each
band) were averaged to get the final photometry. The measurement error
was estimated as the r.m.s. of the individual flux densities.

In the case of the high dynamic range mode observations, we made the
photometry on individual saturation corrected, CBCD, images. Because of
the bright target we used only those 12 short exposure time images
(with integration time of 0.6\,s) where PV~Cep appeared on the
frames. In the course of photometry we used the same method as in the
case of subarray mode observations.

The final uncertainties of the photometry were computed by
adding quadratically the measurement errors and an absolute
calibration error of 2\% (IRAC Data Handbook).

\paragraph*{IRS.} The \textit{Spitzer} spectroscopic observations of PV~Cep were
obtained in spectral mapping mode with the short low
(5.2--14.5$\,\mu$m, R$\approx$60--127), short high (9.9--19.5$\,\mu$m,
R$\approx$600), and long high (18.7--37.2$\,\mu$m, R$\approx$600)
channels. The spectral maps consisted of two nod positions and nine
positions perpendicular to the slit. We only considered the two
central positions, as if it were a normal staring measurement. The data
were processed at the SSC with the S18.7.0 pipeline to BCD level. We
extracted spectra from the two-dimensional dispersed BCD images using
the \textit{Spitzer} IRS Custom Extraction software ({\sevensize SPICE} version 2.1.2). In
the case of the short low channel, we first subtracted the two nod
positions from each other in order to remove the sky background, then
extracted a spectrum from a wavelength-dependent, tapered aperture
around the star. In the case of the high-resolution channels, we
extracted spectra from the full slit. Since no extended emission could
be detected in the IRAC images, we calibrated the spectra by assuming
that PV~Cep is a point source. In case the star was not well centered
in the slit (i.e. the offset from the slit center perpendicular to the
slit was $\ge$0.5\arcsec), we corrected the spectra for flux loss using
the measured IRS beam profiles, following the procedure described in
\citet{Kospal08}. After this, there was still a slight
mismatch between the different channels. The discrepancy could be
eliminated by simply adding a linear correction to the short high
channel.

\paragraph*{MIPS photometry.} 24 and 70$\,\mu$m images of PV~Cep were
obtained as part of the Gould Belt Legacy Survey. We started from BCD
files reduced with the \textit{SSC} pipeline version S16.1. At 70$\,\mu$m, we
used GeRT (version S14.0 v1.1 [060415]) to do column mean subtraction
and time filtering on the BCD files. We used {\sevensize MOPEX} (version 18.1.5) 
to create mosaics at both 24
and 70$\,\mu$m, with pixel scales of 2.45\arcsec \ and 4\arcsec,
respectively. Since PV~Cep was saturated at both wavelengths, aperture
photometry was not possible. In order to determine the flux of PV Cep,
we first obtained aperture photometry for 7 or 5 non-saturated stars
in the field at 24 and 70$\,\mu$m, respectively, (an aperture radius
of 7\arcsec, sky annulus between 40 and 50\arcsec, and aperture correction
of 1.61 was used at 24$\,\mu$m, and an aperture radius of 8\arcsec, sky
annulus between 39 and 65\arcsec, and aperture correction of 3.70 was
used at 70$\,\mu$m). Then we normalized the psf profiles of these
stars to 1\,Jy and averaged them. The obtained psf profile was then
fitted to the non-saturated parts of the profile of PV~Cep and the
scaling factor needed for a good fit directly gave the flux of our
target.

\paragraph*{MIPS spectroscopy.} The low resolution far-IR
(55--95\,$\mu$m, R$\approx$15--25) spectrum of PV~Cep was obtained in
the MIPS spectral energy distribution (SED) mode. Two observing cycles
were performed, each consisting of six pairs of 3\,s long on- and
off-source exposures. The on-source and the background position was
separated by 1\arcmin. The data reduction of the MIPS SED observation was
started with the BCD images (pipeline version S18.12) and the {\sevensize MOPEX}
software was utilized to perform the necessary processing steps
(combination of data, background removal, application of dispersion
solution) and to compile the final image with pixel size of
9.8\arcsec. The spectrum was extracted from the sky-subtracted final
image using a five pixel wide aperture. As a final step an aperture
correction was applied, the aperture correction factors were taken
from \citet{Lu08}.


\clearpage

\begin{table*}
\begin{minipage}{126mm}
\begin{center}
\caption{Comparison stars}
\label{Tab_comp.full}
{\footnotesize
\begin{tabular}{lccccc}
\hline
N & RA(2000) & D(2000) & {\it V}\,($\Delta V$) & {\it R}$_\mathrm{C}$\,($\Delta R_\mathrm{C}$) & {\it I}$_\mathrm{C}$\,($\Delta I_\mathrm{C}$)  \\
 &  h m s & d \ \arcmin \ \arcsec & [mag] & [mag] & [mag] \\
\hline
C1  & 20 46 28.89 & +67 59 05.4 & 14.072\,(0.012) &  13.595\,(0.012) &  13.200\,(0.008) \\
C2  & 20 46 26.50 & +67 58 11.8 & 14.608\,(0.005) &  13.948\,(0.004) &  13.495\,(0.005) \\
C3  & 20 46 10.44 & +67 55 44.2 & 15.134\,(0.005) &  14.205\,(0.004) &  13.637\,(0.004) \\
C4  & 20 46 01.22 & +67 55 26.7 & 16.741\,(0.010) &  15.654\,(0.005) &  14.670\,(0.004) \\
C5  & 20 45 48.42 & +67 59 12.2 & 15.125\,(0.008) &  14.505\,(0.005) &  14.000\,(0.005) \\
C6  & 20 46 37.77 & +67 58 58.4 & 15.648\,(0.008) &  14.907\,(0.005) &  14.289\,(0.005) \\
C7  & 20 45 41.29 & +67 59 28.0 & 16.747\,(0.015) &  16.074\,(0.012) &  15.538\,(0.007) \\
C8  & 20 45 46.21 & +67 55 25.3 & 17.969\,(0.015) &  16.812\,(0.012) &  15.712\,(0.008) \\
C9  & 20 46 22.32 & +67 54 56.2 & 15.676\,(0.010) &  14.938\,(0.008) &  14.363\,(0.007) \\ 
C10 & 20 46 28.89 & +67 55 30.5 & 16.945\,(0.010) &  16.366\,(0.010) &  15.877\,(0.007) \\
C11 & 20 46 36.19 & +67 55 54.7 & 16.210\,(0.010) &  15.739\,(0.010) &  15.345\,(0.007) \\
C12 & 20 46 25.74 & +67 56 54.0 & 16.190\,(0.010) &  15.538\,(0.006) &  15.009\,(0.006) \\
C13 & 20 46 22.09 & +67 59 16.9 & 16.480\,(0.010) &  15.835\,(0.006) &  15.304\,(0.008) \\
C14 & 20 46 19.98 & +67 59 53.4 & 16.163\,(0.010) &  15.631\,(0.005) &  15.212\,(0.005) \\
C15 & 20 46 37.26 & +67 57 41.1 & 14.676\,(0.005) &  14.105\,(0.005) &  13.649\,(0.007) \\ 
C16 & 20 45 43.02 & +67 59 00.1 & 15.723\,(0.009) &  14.995\,(0.005) &  14.412\,(0.007) \\
\hline
\end{tabular}}
\end{center}
\end{minipage}
\end{table*}

\begin{table*}
\begin{minipage}{126mm}
\begin{center}
\caption{Results of the photometry of PV Cep}
\label{Tab_phot.full}
{\footnotesize
\begin{tabular}{rccccl}
\hline
JD & Date &{\it V}\,($\Delta V$) &  {\it R}$_\mathrm{C}$\,($\Delta R_\mathrm{C}$) & {\it I}$_\mathrm{C}$\,($\Delta I_\mathrm{C}$) & Telescope \\
(2454000+) & yyyymmdd & mag & mag & mag  \\
\hline
$-$329 & 20051027 & 14.58\,(0.16)  & 13.28\,(0.06)  & 11.78\,(0.06)  & RCC \\
$-$327 & 20051029 & 15.08\,(0.04)  & 13.39\,(0.04)  & 11.90\,(0.11)  & RCC  \\
$-$326 & 20051030 & 15.43\,(0.06)  & 13.59\,(0.04)  & 11.98\,(0.06)  & RCC  \\
$-$325 & 20051031 & 15.14\,(0.07)  & 13.36\,(0.01)  & 11.84\,(0.16)  & RCC  \\
$-$308 & 20051117 & 15.59\,(0.04)  & 13.76\,(0.05)  & 12.34\,(0.05)  & RCC  \\
1     & 20060922  & 17.46\,(0.05)  & 15.98\,(0.02)  & 14.60\,(0.02)  & RCC  \\  
26    & 20061017  & 17.54\,(0.05)  & 15.98\,(0.02)  & 14.65\,(0.02)  & RCC \\	    
570   & 20080413  & $\cdots$	   & 15.36\,(0.05)  & 13.60\,(0.04)  & Schmidt	\\	    
589   & 20080502  & $\cdots$	   & $\cdots$       & 14.11\,(0.02)  &  Schmidt  \\        
594   & 20080507  & $\cdots$	   & $\cdots$       & 14.25\,(0.04)  &  Schmidt  \\         
626   & 20080608  & $\cdots$	   & $\cdots$       & 14.99\,(0.02)  &  Schmidt  \\         
627   & 20080609  & $\cdots$	   & 16.39\,(0.02)  & 15.03\,(0.02)  & Schmidt  \\	     
645   & 20080627  & $\cdots$	   & 16.55\,(0.02)  & 15.12\,(0.03)  & RCC  \\	
646   & 20080628  & $\cdots$	   & 16.57\,(0.02)  & 15.14\,(0.04)  & RCC  \\	
647   & 20080629  & $\cdots$	   & 16.61\,(0.01)  & 15.16\,(0.01)  & RCC  \\	
649   & 20080701  & $\cdots$	   & 16.70\,(0.01)  & 15.21\,(0.01)  & RCC  \\
650   & 20080702  & $\cdots$	   & 16.61\,(0.03)  & 15.20\,(0.02)  & RCC  \\
675   & 20080727  & $\cdots$	   & 16.74\,(0.04)  & 15.33\,(0.04)  & Schmidt  \\
676   & 20080728  & $\cdots$	   & 16.82\,(0.04)  & 15.43\,(0.04)  & Schmidt  \\
693   & 20080814  & $\cdots$	   & 16.67\,(0.01)  & 15.31\,(0.01)  & Schmidt  \\
705   & 20080826  & 18.37\,(0.03)  & 16.69\,(0.03)  & 15.09\,(0.04)  & CA 2.2-m \\
707   & 20080828  & $\cdots$	   & $\cdots$       & 15.15\,(0.02)  & Schmidt \\	 
708   & 20080829  & 18.04\,(0.03)  & 16.50\,(0.03)  & 15.14\,(0.03)  & CA 2.2-m\\
709   & 20080830  & $\cdots$	   & $\cdots$	    & 15.02\,(0.02)  & Schmidt \\  
710   & 20080831  & $\cdots$	   & 16.35\,(0.02)  & 15.00\,(0.02)  & Schmidt \\	 
738   & 20080928  & 17.80\,(0.02)  & 16.32\,(0.01)  & 15.00\,(0.01)  & RCC   \\
757   & 20081017  & 17.94\,(0.03)  & 16.41\,(0.03)  & 15.07\,(0.03)  & RCC   \\
758   & 20081018  & 17.75\,(0.02)  & 16.29\,(0.01)  & 14.94\,(0.01)  & RCC   \\
760   & 20081020  & 17.81\,(0.02)  & 16.35\,(0.01)  & 15.00\,(0.01)  & RCC   \\
782   & 20081111  & $\cdots$	   & $\cdots$       & 14.94\,(0.01)  & Schmidt \\
808   & 20081207  & $\cdots$	   & 16.24\,(0.04)  & 14.84\,(0.05)  & Schmidt  \\ 
809   & 20081208  & $\cdots$	   & 16.19\,(0.02)  & 14.80\,(0.02)  & Schmidt  \\ 
840   & 20090108  & $\cdots$	   & $\cdots$       & 15.18\,(0.02)  & Schmidt  \\
841   & 20090109  & 18.08\,(0.10)  & 16.54\,(0.07)  & 15.22\,(0.04)  & Schmidt  \\
883   & 20090220  & $\cdots$	   & 16.84\,(0.05)  & 15.41\,(0.06)  & Schmidt  \\  
892   & 20090224  & $\cdots$	   & $\cdots$       & 15.34\,(0.04)  & Schmidt  \\
912   & 20090321  & $\cdots$	   & $\cdots$       & 15.11\,(0.03)  & Schmidt  \\
916   & 20090325  & $\cdots$	   & 16.58\,(0.04)  & 15.24\,(0.04)  & Schmidt  \\
947   & 20090425  & $\cdots$	   & 16.88\,(0.01)  & 15.38\,(0.01)  & Schmidt  \\
949   & 20090427  & $\cdots$	   & 16.74\,(0.01)  & 15.31\,(0.01)  & Schmidt  \\
950   & 20090428  & $\cdots$	   & 16.62\,(0.02)  & 15.24\,(0.01)  & Schmidt  \\
951   & 20090429  & $\cdots$	   & 16.79\,(0.02)  & 15.33\,(0.01)  & Schmidt  \\
953   & 20090501  & $\cdots$	   & 16.79\,(0.02)  & 15.28\,(0.02)  & RCC  \\	     	  
972   & 20090520  & $\cdots$	   & 16.97\,(0.01)  & 15.50\,(0.01)  & RCC  \\
974   & 20090522  & $\cdots$	   & 16.97\,(0.01)  & 15.48\,(0.01)  & Schmidt  \\
976   & 20090524  & $\cdots$	   & 16.35\,(0.03)  & 15.19\,(0.02)  & Schmidt  \\
978   & 20090526  & $\cdots$	   & 16.85\,(0.02)  & 15.43\,(0.02)  & Schmidt  \\
1001  & 20090618  & $\cdots$	   & 16.91\,(0.04)  & 15.44\,(0.05)  & RCC  \\
1022  & 20090709  & $\cdots$	   & 16.91\,(0.04)  & 15.45\,(0.03)  & Schmidt \\
1027  & 20090714  & $\cdots$	   & 16.95\,(0.04)  & 15.49\,(0.04)  & Schmidt \\  
1058  & 20090814  & $\cdots$	   & 17.23\,(0.02)  & 15.70\,(0.02)  & Schmidt \\  
1060  & 20090816  & $\cdots$	   & 17.24\,(0.01)  & 15.70\,(0.02)  & Schmidt \\  
1061  & 20090817  & $\cdots$	   & 17.27\,(0.01)  & 15.71\,(0.02)  & Schmidt \\  
1062  & 20090818  & $\cdots$	   & 17.37\,(0.01)  & 15.86\,(0.02)  & Schmidt \\      
1071  & 20090827  & $\cdots$	   & 17.39\,(0.01)  & 15.78\,(0.01)  & RCC  \\
1072  & 20090828  & $\cdots$	   & 17.40\,(0.01)  & 15.77\,(0.01)  & RCC  \\
\hline
\end{tabular}}
\end{center}
\end{minipage}
\end{table*}

\addtocounter{table}{-1}

\begin{table*}
\begin{minipage}{126mm}
\begin{center}
\caption{Results of the photometry of PV Cep (cont.)}
{\footnotesize
\begin{tabular}{rccccl}
\hline
JD & Date & {\it V}$_\mathrm{C}$\,($\Delta V_\mathrm{C}$) & {\it R}$_\mathrm{C}$\,($\Delta R_\mathrm{C}$) & {\it I}\,($\Delta I$) & Telescope \\
(2454000+) & yyyymmdd & mag & mag & mag  \\
\hline
1077  & 20090902 & $\cdots$	  &  $\cdots$      & 15.48\,(0.01) & RCC  \\	 
1093  & 20090918 & $\cdots$	  &  16.95\,(0.02) & 15.21\,(0.02) & RCC  \\	
1094  & 20090919 & $\cdots$	  &  16.93\,(0.02) & 15.10\,(0.02) & RCC  \\	
1095  & 20090920 & $\cdots$	  &  16.87\,(0.02) & 15.17\,(0.02) & RCC  \\	
1096  & 20090921 & $\cdots$	  &  16.91\,(0.02) & 15.17\,(0.02) & RCC  \\	
1101  & 20090926 & 18.24\,(0.06)  &  16.59\,(0.01) & 14.93\,(0.01) & RCC  \\
1103  & 20090928 & $\cdots$	  &  16.46\,(0.02) & 14.73\,(0.01) & Schmidt  \\	 
1113  & 20091008 & $\cdots$	  &  17.05\,(0.03) & 15.35\,(0.03) & Schmidt  \\	
1114  & 20091009 & 18.69\,(0.04)  &  16.95\,(0.01) & 15.32\,(0.01) & CA 2.2-m \\	
1115  & 20091010 & $\cdots$	  &  16.96\,(0.04) & 15.48\,(0.06) & Schmidt  \\	
1116  & 20091011 & 18.70\,(0.04)  &  17.00\,(0.03) & 15.33\,(0.03) & CA 2.2-m \\	
1119  & 20091014 & 18.60\,(0.01)  &  16.83\,(0.01) & 15.12\,(0.01) & CA 2.2-m \\	
1130  & 20091025 & $\cdots$	  &  16.85\,(0.02) & 15.08\,(0.02) & IAC-80  \\ 
1137  & 20091101 & 18.18\,(0.04)  &  16.61\,(0.02) & 14.66\,(0.01) & IAC-80  \\   
1139  & 20091103 & $\cdots$	  &  16.54\,(0.02) & 14.59\,(0.02) & IAC-80  \\   
1140  & 20091104 & $\cdots$	  &  16.59\,(0.02) & 14.73\,(0.02) & IAC-80 \\   
1142  & 20091106 & 18.21\,(0.04)  &  16.39\,(0.02) & 14.46\,(0.02) & IAC-80 \\   
1143  & 20091107 & 18.08\,(0.03)  &  16.40\,(0.02) & 14.45\,(0.02) & IAC-80 \\   
1149  & 20091113 & $\cdots$	  &  16.51\,(0.03) & 14.58\,(0.04) & Schmidt  \\  
1155  & 20091119 & $\cdots$	  &  16.95\,(0.04) & 15.33\,(0.04) & Schmidt  \\ 
1156  & 20091120 & $\cdots$	  &  16.72\,(0.03) & 15.12\,(0.03) & Schmidt  \\  
1159  & 20091123 & $\cdots$	  &  16.95\,(0.01) & 15.35\,(0.02) & Schmidt  \\  
1161  & 20091125 & $\cdots$	  &  16.98\,(0.01) & 15.32\,(0.01) & Schmidt  \\  
1212  & 20100115 & $\cdots$	  &  16.45\,(0.05) & 14.53\,(0.06) & Schmidt  \\  
1213  & 20100116 & $\cdots$	  &  16.37\,(0.06) & 14.46\,(0.07) & Schmidt  \\  
1253  & 20100225 & $\cdots$	  &  16.27\,(0.13) & 14.56\,(0.12) & Schmidt  \\     
1259  & 20100303 & $\cdots$	  &  16.36\,(0.13) & 14.38\,(0.12) & Schmidt  \\     
1288  & 20100401 & $\cdots$	  &  15.55\,(0.07) & 13.76\,(0.05) & Schmidt  \\
1295  & 20100408 & $\cdots$	  &  15.87\,(0.01) & 13.83\,(0.05) & RCC  \\
1331  & 20100514 & $\cdots$	  &  15.50\,(0.02) & 13.56\,(0.02) & RCC  \\
1372  & 20100624  & 17.70\,(0.06) &  15.81\,(0.03) & 14.02\,(0.03) & Schmidt  \\	
1420  & 20100812  & 17.45\,(0.01) &  15.53\,(0.01) & 13.64\,(0.01) & RCC  \\	
1435  & 20100826  &  $\cdots$     &  16.19\,(0.04) & 14.31\,(0.05) & Schmidt  \\   
1461  & 20100921 & $\cdots$	  &  15.58\,(0.03) & 13.53\,(0.03) & RCC  \\  
1463  & 20100923 & $\cdots$	  &  15.90\,(0.05) & 14.06\,(0.04) & Schmidt  \\  
1464  & 20100924 & $\cdots$	  &  15.86\,(0.05) & 13.99\,(0.06) & Schmidt  \\  
1477  & 20101007 & $\cdots$	  &  $\cdots$	   & 14.13\,(0.03) & Schmidt  \\ 
1524  & 20101123 & $\cdots$	  &  16.15\,(0.04) & 14.45\,(0.04) & Schmidt  \\ 
\hline			 		 	 
\end{tabular}}		 
\end{center}		 
\end{minipage}		 
\end{table*}

\begin{table*} 
\caption{\textit{Spitzer} photometry for PV Cep. All fluxes are color-corrected and given in Jy.}
\label{Tab_Spitzer.full}
\centering
{\small
\begin{tabular}{c c c c c c c}
\hline
\hline
Date        & F$_{3.6\mu\rm m}$ & F$_{4.5\mu\rm m}$ & F$_{5.7\mu\rm m}$ & F$_{8.0\mu\rm m}$ & F$_{24\mu\rm m}$ & F$_{70\mu\rm m}$ \\
\hline
20041029 & 4.47\,(0.09) & 6.98\,(0.14) & 8.34\,(0.17) & 11.85\,(0.23) &             &            \\
20061126 & 1.84\,(0.04) & 3.99\,(0.08) & 5.45\,(0.11) & 8.16\,(0.17)  &             &            \\
20070226 &            &            &            &             & 27.12\,(1.04) & 35.22\,(2.13)\\
20090916 & 3.54\,(0.03) & 5.97\,(0.02) \\
20090917 & 3.16\,(0.03) & 5.41\,(0.03) \\
20090918 & 3.19\,(0.07) & 5.31\,(0.03) \\
20090919 & 3.31\,(0.03) & 5.48\,(0.03) \\
20090920 & 3.22\,(0.02) & 5.34\,(0.02) \\
20090921 & 3.11\,(0.04) & 5.19\,(0.04) \\
20090922 & 3.29\,(0.04) & 5.40\,(0.03) \\
20100109 & 3.27\,(0.07) & 5.53\,(0.03) \\
20100110 & 3.14\,(0.04) & 5.31\,(0.02) \\
20100111 & 3.31\,(0.05) & 5.49\,(0.02) \\
20100112 & 3.34\,(0.08) & 5.67\,(0.02) \\
20100113 & 3.26\,(0.04) & 5.46\,(0.03) \\
20100114 & 3.00\,(0.08) & 5.16\,(0.02) \\
20100115 & 3.05\,(0.02) & 5.23\,(0.02) \\
\hline
\end{tabular}}
\end{table*}

\end{document}